\documentclass[twoside]{dis08}
\usepackage[latin1]{inputenc}
\usepackage[dvips]{graphicx,epsfig,color}
\usepackage{wrapfig,rotating}
\usepackage{amssymb,amsmath,array}
\usepackage{mathrsfs}

\newcommand{\bea}{\begin{eqnarray}}
\newcommand{\eea}{\end{eqnarray}}
\newcommand{\mA}{\mathscr{A}_{TT}}

\begin{document}
\title{Double Transverse-Spin Asymmetries for Drell-Yan Process in $pp$ and $p\bar{p}$ Collisions: 
Role of Nonleading QCD Corrections at Small Transverse Momentum
}

\author{Hiroyuki Kawamura$^1$ and Kazuhiro Tanaka$^2$
%
%
\vspace{.3cm}\\
%
1- Department of Mathematical Sciences,
   University of Liverpool,\\
   Liverpool L69 3BX, United Kingdom
%
\vspace{.1cm}\\
2- Department of Physics, Juntendo University, Inba, Chiba 270-1695, Japan}

\maketitle

\begin{abstract}
We discuss
the double-spin asymmetries 
in transversely polarized Drell-Yan process, calculating 
all-order gluon resummation corrections up to the next-to-leading logarithmic accuracy.
This resummation is 
relevant when the transverse-momentum $Q_T$
of the produced lepton pair is small, and 
reproduces the (fixed-order) next-to-leading QCD corrections upon integrating over $Q_T$.
The resummation corrections behave 
differently between $pp$- and $p\bar{p}$-collision cases
and 
are
small for the latter case at the kinematics 
in the proposed GSI experiments.
This fact allows us to predict large value of 
the double-spin asymmetries 
at GSI,
using the recent
empirical information on the transversity.
\end{abstract}


The double-spin asymmetry in Drell-Yan process 
with transversely-polarized protons,
$p^{\uparrow}p^{\uparrow}
\rightarrow l^+l^-X$, for azimuthal angle $\phi$ of a lepton
measured in the rest frame of the dilepton $l^+l^-$ with invariant mass $Q$ and rapidity $y$,
is given by ($d\omega \equiv dQ^2 dy d\phi$, $q=u,\bar{u},d,\bar{d}, \ldots$)
\begin{equation}
A_{TT}  = 
\frac{d\sigma^{\uparrow  \uparrow}/d\omega - d\sigma^{\uparrow  \downarrow}/d\omega}
{d\sigma^{\uparrow  \uparrow}/d\omega + d\sigma^{\uparrow  \downarrow}/d\omega}
 \equiv \frac{\Delta _T d\sigma /d\omega}{d\sigma /d\omega}
= \frac{\cos (2\phi )}{2}\frac{\sum_q e_q^2 \delta q(x_1 , Q^2)\delta \bar q(x_2 , Q^2) +\cdots}
{\sum_q e_q^2 q(x_1 ,Q^2)\bar q(x_2 ,Q^2) +\cdots}\ ,
\label{eq:1}
\end{equation}
as 
the ratio of products of the relevant 
quark and antiquark 
distributions,
the transversity $\delta q(x, Q^2)$ and the unpolarized $q(x, Q^2)$, and
the ellipses stand for the corrections of next-to-leading order (NLO, $O(\alpha_s)$) and higher
in QCD perturbation theory.
The scaling variables $x_{1,2}$ 
represent the momentum fractions associated with the partons
annihilating via the Drell-Yan mechanism, such that 
$Q^2=(x_1 P_1 + x_2 P_2)^2 = x_1 x_2 S$ and $y=( 1/2)\ln (x_1 /x_2)$,
where 
$S=(P_1 +P_2)^2$ is the CM energy squared of the colliding protons.
Thus the transversely polarized Drell-Yan (tDY) data for (\ref{eq:1}) can provide a direct access 
to the transversity,
and it is important to clarify the role of QCD corrections in the double-spin asymmetries.

It has been shown that the NLO QCD corrections for (\ref{eq:1})
are not so significant
and the resulting $A_{TT}$ is less than a few percent at RHIC, similarly to the LO estimates
(see \cite{KKST:06}).
This reflects that the sea-quark region is probed at RHIC for $Q^2 \gtrsim 10$ GeV$^2$,
where the denominator in (\ref{eq:1}) is enhanced with small $x_{1,2}$.
Now, when the transverse momentum $Q_T$ of the final dilepton is also observed in tDY,
we obtain the double-spin asymmetry at a measured $Q_T$,
as the ratio of the $Q_T$-differential cross sections, 
$\mA (Q_T) = (\Delta _T d\sigma/d\omega dQ_T)/(d\sigma /d\omega dQ_T)$.
In principle,
the relevant parton distributions in this asymmetry
may be controlled by 
the new scale 
$\sim Q_T$,
in contrast to $Q$ in (\ref{eq:1}). 
The small-$Q_T$ case
is important because the bulk of events is produced for $Q_T \ll Q$.
In this case, the cross sections $(\Delta _T ) d\sigma/d\omega dQ_T$ 
receive the large perturbative corrections 
with logarithms $\ln ( Q^2 / Q_T^2 )$ multiplying $\alpha_s$
at each order by the recoil 
from gluon radiations, which have to be resummed
to all orders \cite{KKST:06}. 
As a result, 
we get ($b_0\equiv 2e^{-\gamma_E}$ with $\gamma_E$ the Euler constant)
\begin{equation}
\mA(Q_T) =\frac{\cos (2\phi )}{2}
\frac{\int d^2 b\ e^{i \mathbf{b} \cdot \mathbf{Q}_T} 
e^{S(b,Q)} 
\sum\nolimits_q  e_q^2 
\delta q (x_1 , b_0^2/b^2 ) 
\delta \bar{q}(x_2 , b_0^2/b^2 ) +\cdots}
{\int d^2 b\ e^{i \mathbf{b} \cdot \mathbf{Q}_T} 
e^{S(b,Q)} 
 \sum\nolimits_q  e_q^2 
q (x_1 , b_0^2/b^2 ) 
\bar{q}(x_2 ,b_0^2/b^2 ) +\cdots
},
\label{eq:2}
\end{equation}
where 
the numerator and denominator are, respectively, reorganized
in the impact parameter $b$ space in terms of 
the Sudakov factor $e^{S(b,Q)}$ resumming soft and flavor-conserving collinear
radiation, while the ellipses involve the remaining contributions of the $O (\alpha_s)$
collinear radiation, which can be absorbed into the exhibited terms
as 
$\delta q \rightarrow \Delta_T C_{qq}\otimes  \delta q$,
$q \rightarrow C_{qq} \otimes  q + C_{qg}\otimes g$ using the corresponding coefficient 
functions $(\Delta_T) C_{ij}$; there appears no gluon distribution in the numerator
of (\ref{eq:2}), similarly as in (\ref{eq:1}), 
because of the chiral-odd nature.
Using 
{\em universal} Sudakov exponent $S(b,Q)$ with the first nonleading 
anomalous dimensions in (\ref{eq:2}), 
the first three towers of large logarithmic contributions
to the cross sections,
$\alpha_s^n\ln^m(Q^2/Q_T^2)/Q_T^2$ ($m=2n-1, 2n-2, 2n-3$),
are resummed to all orders in $\alpha_s$,
yielding the next-to-leading logarithmic (NLL) resummation.
In addition to these NLL resummed components relevant for small $Q_T$,
the ellipses in (\ref{eq:2}) also involve  
the other terms of the fixed-order $\alpha_s$, which treats the LO processes in the large
$Q_T$ region, so that (\ref{eq:2}) 
is the ratio of the NLL+LO polarized and unpolarized cross sections.
We include a Gaussian smearing 
as usually
as $S(b,Q) \rightarrow S(b,Q) -g_{NP}b^2$,
corresponding to intrinsic transverse momentum.
The integration of these NLL+LO cross sections 
$\Delta_T  d\sigma/d\omega dQ_T$, $d\sigma/d\omega dQ_T$ over $Q_T$
coincides \cite{KKST:06} with 
the NLO cross sections 
$\Delta_T d\sigma/d\omega$,
$d\sigma/d\omega$,
respectively,
associated with $A_{TT}$ of~(\ref{eq:1});
thus the NLO parton distributions have to be substituted into 
(\ref{eq:2}) as well as 
(\ref{eq:1}).

The resummation indeed 
makes $1/b \sim Q_T$ the relevant scale.
The numerical evaluation of (\ref{eq:2})
at NLL+LO
with RHIC and J-PARC kinematics has revealed \cite{KKST:06}
that, in small and moderate $Q_T$ region ($Q_T \lesssim Q$), 
$\mA(Q_T)$ is governed by the NLL resummed component
and is almost constant as a function of $Q_T$,
reflecting universality of the large Sudakov effects.
The results show $\mA(Q_T)> A_{TT}$, because the denominator of (\ref{eq:2})
is not enhanced for $Q_T \ll Q$ compared with 
that of the corresponding NLO $A_{TT}$ of (\ref{eq:1}),
and also show the tendency
that $\mA(Q_T)$ with resummation at higher level yields the larger value.
Using the NLO transversities 
that saturate the Soffer bound, $2\delta q(x,\mu^2)\le q(x,\mu^2)+\Delta q(x,\mu^2)$, 
at a low scale $\mu$ with $\Delta q$ the helicity distribution, 
the NLO value of (\ref{eq:1}) at $\phi=0$ is 
$\lesssim 4$\% and $\sim 13$\% for typical kinematics
at RHIC and J-PARC, respectively, and 
the NLL+LO
$\mA(Q_T)$ for small $Q_T$ using the same transversity 
are larger 
than those 
NLO $A_{TT}$
by about 20-30\%~\cite{KKST:06}.
It is also worth noting that, for $Q_T \approx 0$,
the $b$ integral of (\ref{eq:2})
is controlled by a saddle point $b=b_{SP}$, which has the same
value between the numerator and denominator in (\ref{eq:2}) 
at NLL accuracy \cite{KKST:06}:
combined
with the almost constant behavior of $\mA(Q_T)$ mentioned above, 
\begin{equation}
\mA(Q_T) \simeq \mA(0)\simeq \frac{\cos (2\phi )}{2}\frac{\sum_q e_q^2 \delta q(x_1 , b_0^2/b_{SP}^2 )
\delta \bar q(x_2 , b_0^2/b_{SP}^2 )}
{\sum_q e_q^2 q(x_1 ,b_0^2/b_{SP}^2 )\bar q(x_2 ,b_0^2/b_{SP}^2 )}\ ,
\label{eq:3}
\end{equation}
for small $Q_T$ region, omitting the small corrections
from the LO components involved in the ellipses in (\ref{eq:2}).
The saddle-point evaluation 
does not lose the NLL accuracy of (\ref{eq:2}); in particular, the $O(\alpha_s)$ contributions
from the coefficients $(\Delta_T) C_{ij}$, e.g. those with gluon distribution
in the denominator, completely decouple as $Q_T \rightarrow 0$.
Remarkably \cite{KKST:06}, $b_0/b_{SP}\simeq 1$ GeV, 
irrespective of the values of $Q$ and $g_{NP}$.
The 
formula (\ref{eq:3}) 
allows quantitative evaluation of (\ref{eq:2})
to good accuracy, 
and embodies the above features 
of $\mA(Q_T)$
in a compact form.
Next we discuss 
the $p\bar{p}$-collision case,
$p^{\uparrow}\bar{p}^{\uparrow}
\rightarrow l^+l^-X$; here and below,
the formal interchange, $\delta q(x_2 ) \leftrightarrow \delta \bar{q}(x_2 )$,
$q(x_2 ) \leftrightarrow \bar{q}(x_2 )$, for the distributions 
associated with the variable $x_2$ should be understood in 
the relevant formulae (\ref{eq:1})-(\ref{eq:3}) 
for the asymmetries.
Thus this case allows us to probe the product of the two quark-transversities, 
in particular,
the valence-quark transversities
for the region $0.2 \lesssim x_{1,2} \lesssim 0.7$ 
in the proposed polarization experiments at GSI (see e.g. \cite{SSVY:05}).
When the transverse-momentum $Q_T$ is unobserved, 
one obtains $A_{TT}$ of (\ref{eq:1}): for this asymmetry 
at GSI,
the NLO ($O(\alpha_s)$) corrections as well as the
higher order corrections beyond them in the framework of the threshold resummation
are shown to be 
rather small, so that the LO value of 
$A_{TT}$, which turns out to be large, is rather robust \cite{SSVY:05}. 

We now consider the QCD corrections
at a measured $Q_T$, calculating
$\mA (Q_T)$ of (\ref{eq:2}), (\ref{eq:3})
at GSI kinematics.
The numerical evaluation of (\ref{eq:2}) 
using the transversity distributions
corresponding to the Soffer bound, 
which are same as in the $pp$-collision case discussed above,
shows \cite{KKT:08} that the NLL resummed component dominates $\mA(Q_T)$ 
in small and moderate $Q_T$ region such that $\mA(Q_T)$ is almost constant, with 
even flatter behavior than for the $pp$ case.
It is also demonstrated that $\mA(Q_T)$ at NLL+LO has almost the 
same value as that at LL; i.e., in contrast to the $pp$ case,
the resummation at higher level does not 
enhance the asymmetry.
We here note that $\mA(Q_T)$ at LL is given by (\ref{eq:2}) 
omitting all nonleading corrections, i.e., omitting the ellipses,
replacing $S(b,Q)$ by that at the LL level, and replacing the scale of the parton distributions 
as $b_0^2/b^2 \rightarrow Q^2$,
so that the result coincides with $A_{TT}$ of (\ref{eq:1}) at LO.
Combined with the above-mentioned property of $A_{TT}$,
we obtain, for $Q_T \lesssim Q$,
\begin{equation}
\mA(Q_T) \simeq A_{TT}\ ,
\label{eq:4}
\end{equation}
at GSI, with the large value 
of the asymmetry which is quite stable when including the QCD
(resummation and fixed-order) corrections.

To clarify the reason behind this remarkable difference between the
$p\bar{p}$- and $pp$-collision cases, the saddle-point formula 
(\ref{eq:3}) is useful.
The simple form of (\ref{eq:3}) is reminiscent of $A_{TT}$ of (\ref{eq:1}) at LO,
but is different from the latter, only in the 
unconventional scale $b_0^2/b_{SP}^2$.
In fact, this scale, $b_0^2/b_{SP}^2 \simeq 1$ GeV$^2$ ($\ll Q^2$) at all GSI kinematics 
as determined by the saddle point,
completely absorbs the nonuniversal effects
associated with nonleading (NLL) level resummation, because $\mA^{\rm LL}(Q_T)=A_{TT}^{\rm LO}$
as noted above.
In the valence region $0.2 \lesssim x_{1,2} \lesssim 0.7$ 
relevant for GSI kinematics, the $u$-quark contribution dominates in (\ref{eq:3})
and (\ref{eq:2}), so that these asymmetries are controlled by the ratio 
of the $u$ quark distributions,
$\delta u(x_{1,2}, \mu^2)/u(x_{1,2},\mu^2)$,
with $\mu^2 = b_0^2/b_{SP}^2$ and $Q^2$, 
respectively.
It is straightforward to see that 
the scale dependence in this ratio
almost cancels between the numerator and denominator 
in the valence region
as $\delta u(x,  b_0^2/b_{SP}^2)/u(x,  b_0^2/b_{SP}^2)
\simeq \delta u(x, Q^2)/u(x, Q^2)$
(see Fig.~3 in \cite{KKT:08}), implying (\ref{eq:4})
at GSI;
this is not the case 
for $pp$ collisions at RHIC and J-PARC,
because of very different behavior of the sea-quark components under the evolution
between transversity and unpolarized distributions~\cite{KKST:06}.
A similar logic applied to (\ref{eq:2})
also explains why $\mA(Q_T)$
in $p\bar{p}$ collisions at GSI
are flatter 
than in $pp$ collisions as mentioned above.

\begin{figure}[!t]
\begin{center}
\hspace*{-0.5cm}
\includegraphics[width=0.55\textheight,clip]{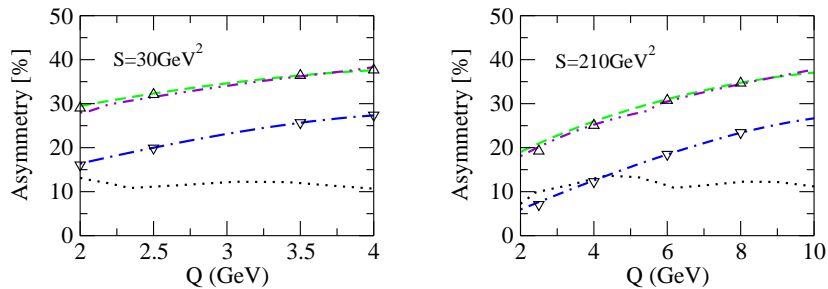}
\end{center}
\caption{The double transverse-spin asymmetries at GSI as functions of $Q$ 
with $y=\phi=0$.}
\end{figure}
Another consequence of the similar logic 
is that
$\delta u(x, 1\ {\rm GeV}^2)/u(x, 1\ {\rm GeV}^2)$ 
as a function of $x$ directly determines the $Q$- as well as $S$-dependence 
of the value of (\ref{eq:4}) at GSI, 
with $x_{1,2}=(Q/\sqrt{S})e^{\pm y}$. 
In Fig.~1, using the NLO transversity distributions
corresponding to the Soffer bound,
the symbols ``{\scriptsize $\bigtriangleup$}''
plot $\mA(Q_T)$
of (\ref{eq:2}) at NLL+LO 
as a function of 
$Q$ with $y=\phi=0$ and $Q_T \simeq 1$ GeV,
in the fixed-target ($S=30$ GeV$^2$) 
and collider ($S=210$ GeV$^2$) modes at GSI~\cite{KKT:08}.
The dashed curve draws the result using 
(\ref{eq:3});
this simple 
formula indeed works well. 
Also plotted by the two-dot-dashed curve
is $A_{TT}$ of (\ref{eq:1}) at LO
with the transversities corresponding to the Soffer bound at LO level,
to demonstrate (\ref{eq:4}).
The $Q$- and $S$-dependence of these results 
reflects that
the ratio $\delta u(x, 1 {\rm GeV}^2)/u(x, 1 {\rm GeV}^2)$ 
is an increasing function 
of $x$ for the present choice.
These results using the Soffer bound 
show the ``maximally possible'' asymmetry, i.e., 
optimistic 
estimate.
A more realistic estimate of (\ref{eq:2}) and (\ref{eq:3}) 
is shown \cite{KKT:08} in Fig.~1 by the symbols
``{\scriptsize $\bigtriangledown$}'' and the dot-dashed curve, respectively,
with the NLO transversity distributions
assuming $\delta q(x, \mu^2)=\Delta q(x, \mu^2)$ at a low scale $\mu$,
as suggested by various
nucleon models and favored
by the results of empirical fit for transversity \cite{Anselmino:07}.
The new estimate gives 
smaller asymmetries
compared with the Soffer bound results
because the $u$-quark transversity is considerably smaller,
but still yields rather large asymmetries \cite{KKT:08}.
Based on (\ref{eq:4}), these results also give estimate of $A_{TT}$ of (\ref{eq:1}).


At present, empirical information of transversity is based on
the LO global fit, using
the semi-inclusive deep inelastic scattering data and 
assuming that the antiquark transversities in the proton vanish, $\delta \bar{q}(x)=0$,
so that the corresponding LO
parameterization is
available only for $u$ and $d$ quarks \cite{Anselmino:07}.
Fortunately, however, the dominance of the $u$-quark contribution
in the GSI kinematics allows quantitative evaluation of $A_{TT}$ at LO
using only this empirical information \cite{KKT:08}:
the upper limit of the one-sigma error bounds for the $u$- and $d$-quark transversities
obtained by the global fit \cite{Anselmino:07}
yields the ``upper bound'' of $A_{TT}$ shown
by the dotted curve in Fig.~1. 
Using (\ref{eq:4}), this result would also represent estimate 
of $\mA(Q_T)$.
In the small $Q$ region, our full NLL+LO result of $\mA(Q_T)$,
shown by ``{\scriptsize $\bigtriangledown$}'',  can be consistent
with estimate using the empirical LO transversity, but these results have rather 
different behavior
for increasing $Q$, 
because the $u$-quark transversity for the former
lies, for $x \gtrsim 0.3$, slightly outside the one-sigma error bounds of the global 
fit \cite{KKT:08}.
Thus, 
the asymmetries to be observed at GSI, 
in particular, the behavior 
of $\mA(Q_T)$ as well as $A_{TT}$
as functions of $Q$, 
will allow us to determine
the detailed shape of transversity distributions.
Other interesting DY spin-asymmetries at GSI are
the longitudinal-transverse asymmetry $A_{LT}$ \cite{KYT:08} 
and the single transverse-spin asymmetry \cite{JQWY:06}, 
which are sensitive to twist-3 effects inside proton.

\vspace{0.1cm}
This work was supported by the Grant-in-Aid for Scientific Research 
No.~B-19340063. 

\begin{footnotesize}



%

\end{footnotesize}


\end{document}